\documentclass[11pt]{article}

\usepackage[top=0.7in, bottom=1in, left=1in, right=1in]{geometry}
\usepackage{amsmath, amssymb, amsthm}
\usepackage{graphicx}
\usepackage{natbib}
\usepackage{hyperref}
\usepackage{setspace}
\usepackage{authblk}
\usepackage{bm}
\usepackage[ruled]{algorithm2e}
\usepackage{algpseudocode}
\usepackage{booktabs}
\usepackage{multirow} 
\usepackage{subfigure}
\title{Transportability of Regression Calibration with External Validation Studies for Measurement Error Correction}
\author[1]{Zexiang Li}
\author[1,2]{Donna Spiegelman}
\author[3,4]{Molin Wang}
\author[1,5]{Zuoheng Wang}
\author[1,2]{Xin Zhou}
\affil[1]{Department of Biostatistics, Yale School of Public Health, New Haven, CT}
\affil[2]{Center on Methods for Implementation and Prevention Science, Yale School of Public Health, New Haven, CT}
\affil[3]{Department of Epidemiology, Harvard T.H. Chan School of Public Health, Boston, MA}
\affil[4]{Department of Biostatistics, Harvard T.H. Chan School of Public Health, Boston, MA}
\affil[5]{Department of Biomedical Informatics \& Data Science, Yale School of Medicine, New Haven, CT}
\date{}

\makeatletter
\def\bSig\mathbf{\Sigma}

\def\T{{ \mathrm{\scriptscriptstyle T} }}

\begin{document}

\maketitle

\begin{abstract}

In nutritional and environmental epidemiology, exposures are impractical to measure accurately, while practical measures for these exposures are often subject to substantial measurement error. Regression calibration is among the most used measurement error correction methods with external validation studies. The use of external studies to assess the measurement error process always carries the risk of introducing estimation bias into the main study analysis. Although the transportability of regression calibration is usually assumed for practical epidemiology studies, it has not been well studied. In this work, under the measurement error process with a mixture of Berkson-like and classical-like errors, we investigate conditions under which the effect estimate from regression calibration with an external validation study is unbiased for the association between exposure and health outcome. We further examine departures from the transportability assumption, under which the regression calibration estimator is itself biased. However, we theoretically prove that, in most cases, it yields lower bias than the naive method. The derived conditions are confirmed through simulation studies and further verified in an example investigating the association between the risk of cardiovascular disease and moderate physical activity in the health professional follow-up study. 

\end{abstract}

\section{Introduction}

Exposures of interest in epidemiologic research, such as dietary intake and physical activity, are frequently measured with error \citep{Thiebaut2007, Ferrari2007}. When such errors are not properly accounted for, the estimates of exposure-outcome association may be biased \citep{buzasMeasurementError2014}. Regression calibration (RC) \citep{carrollMeasurementErrorNonlinear2006, rosnerCorrectionLogisticRegression1989} is one of the most widely applied approaches to address this problem \citep{Shaw2018}. RC typically involves two models: a measurement error model (MEM), which regresses the reference measurement for true exposure on the surrogate exposure and confounders, and an outcome model, which relates the surrogate exposure or the predicted true exposure to the outcome without measurement error. The MEM parameters are usually estimated from a validation study, in which both the surrogate and reference measurements for true exposures are observed. The estimated MEM is then applied to the main study, where only the surrogate exposure is available, to adjust for measurement error in the outcome model. 

In practice, conducting an internal validation study within the main study is often costly. When such a study design is infeasible, an external validation study (EVS), which is independent of the main study, is commonly employed. For example,  the Men’s and Women’s Lifestyle Validation Studies \citep{MLVS, WLVS} were designed to assess measurement error in self-reported diet and physical activity and can serve as EVS. Reliance on an EVS may nonetheless lead to biased estimates if the transportability assumption required by RC, that the MEM derived from the EVS accurately represents that of the main study, is violated \citep{carrollMeasurementErrorNonlinear2006}. To address this issue, \cite{li2025} proposed a modified RC approach that accommodates differences between the main study and the EVS. When the transportability assumption does hold, however, this alternative estimator is less efficient than the standard RC. Assessing the validity of the transportability assumption is therefore essential. Although \cite{li2025} suggested using the conditional distribution of the surrogate exposure given confounders as an indicator, this provides only a sufficient condition instead of a necessary one, as shown in Sections 2 and 3. Relying solely on this criterion may incorrectly classify some transportable situations as violations, leading to unnecessary loss of efficiency.  

Another practical concern is that the gold standard for the true exposure is often unavailable. In such cases, reference measurements obtained from validation studies can only approximate true exposures and may still contain random error. This type of measurement is commonly referred to as an alloyed gold standard exposure \citep{Spiegelman1997}. When the transportability assumption linking the main study and the EVS is violated in this scenario, no specific method has been proposed, and RC remains the method commonly used to correct for measurement error. However, because RC is biased under such violations, a natural question, then, is the extent to which violations of the transportability assumption affect the relative unbiasedness of the RC estimator compared with the naive method. Specifically, the naive method only depends on the main study, whereas RC draws information from both the main study and the EVS. When the situations in these two studies differ, it is unclear which method yields the estimate with a smaller bias for the main study analysis.

Motivated by these considerations, this work aims to address the question of the transportability assumption in settings where the true exposures are unobserved within the RC framework.  We specifically investigate (1) the necessary conditions that guarantee the validity of the RC estimator, and (2) how violations of the transportability assumption influence the relative bias of the RC estimator compared with the naive approach.

The remainder of the paper is organized as follows. Section 2 discusses the necessary conditions for an unbiased and valid RC estimator, as well as the scenarios under which the RC method exhibits smaller bias than the naive estimate. In Section 3, we design a series of simulation studies to verify the theoretical findings from Section 2. Section 4 demonstrates our conclusions using a practical application. Finally, Section 5 concludes with a discussion. 

\section{Methods}

Consider a main study and an EVS with sample sizes $n_m$ and $n_v$, respectively. Let $\bm{Y}$ be an outcome, $\bm{x}$ be $p$-dimensional unobserved true exposures, $\bm{X}$ be $p$-dimensional reference measurements for true exposures, $\bm{Z}$ be $p$-dimensional surrogate exposures, and $\bm{W}$ be $q$-dimensional error-free confounders. In the main study, data are available for each subject $i$ as $(Y_i, \bm{Z}_i, \bm{W}_i)$, where $i = 1, \ldots, n_m$. In the EVS, we observe $(\bm{X}_i, \bm{Z}_i, \bm{W}_i)$ for each subject $i = n_m + 1, \ldots, n_m + n_v $. 

\subsection{Review}

In the absence of measurement error, we assume that, given $\bm{x}_i$ and $\bm{W}_i$, the outcome $Y_i$ follows an exponential family distribution, $f(y_i) = \exp\{(y_i\eta_i-b(\eta_i))/a(\phi) + c(y_i, \phi)\}$\citep{mccullaghGeneralizedLinearModels1989}, with link function $g$ for which
	\begin{equation*}
		g\{E(Y_i|\bm{x}_i,\bm{W}_i)\} = \beta_0 + \bm{\beta}_1^\T \bm{x}_i + \bm{\beta}_2^\T \bm{W}_i,
	\end{equation*}
where $\eta_i = (b')^{-1}\{g^{-1}(\beta_0 + \bm{\beta}_1^\T \bm{x}_i + \bm{\beta}_2^\T \bm{W}_i)\}$, $\beta_0$ represents the intercept, $\bm{\beta}_1$ and $\bm{\beta}_2$ are regression coefficient vectors of dimensions $p$ and $q$, respectively. The primary parameter of interest is the exposure-outcome association $\bm{\beta}_1$. 

However, in practice, $\bm{x}_i$ is not observed in the main study due to measurement error, and thus $\bm{\beta}_1$ cannot be directly estimated. 
Instead, using the surrogate exposures $\bm{Z}_i$, we fit
\begin{equation}\label{eq:outcome2}
		g\{\mathbb{E}(Y_i|\bm{Z}_i,\bm{W}_i)\} = \beta_0^* + {\bm{\beta}_1^{*}}^{\T} \bm{Z}_i + {\bm{\beta}_2^{*}}^{\T} \bm{W}_i,
\end{equation}
where $\eta_i = (b')^{-1}\{g^{-1}(\beta_0^* + {\bm{\beta}_1^{*}}^{\T} \bm{Z}_i + {\bm{\beta}_2^{*}}^{\T} \bm{W}_i)\}$, $\beta_0^*$ represents the intercept, $\bm{\beta}_1^*$ and $\bm{\beta}_2^*$ are regression coefficient vectors of dimensions $p$ and $q$, respectively. The estimate $\widehat{\bm{\beta}}_1^*$, known as the naive estimator of $\bm{\beta}_1$, is well established to be biased due to measurement error. 

To account for measurement error, the MEM is specified in the validation study as
\begin{equation}
	E(\bm{X}_i\mid \bm{Z}_i, \bm{W}_i) = \bm{\gamma}_{0} + \bm{\Gamma}_{1}^\T \bm{Z}_i + \bm{\Gamma}_{2}^\T \bm{W}_i, \label{eq:MEM}
\end{equation}
where $\bm{\gamma}_0$ is a $p$-dimensional vector, $\bm{\Gamma}_1$ and $\bm{\Gamma}_2$ are $p\times p$ and $q\times p$ matrices, respectively. When the random error in $\bm{X}$ is uncorrelated with the measurement error in $\bm{Z}$, an assumption routinely made in applications \citep{MOTEX1, MOTEX2, MOTEX3}, \cite{Spiegelman1997} showed that $E(\bm{x}_i\mid \bm{Z}_i, \bm{W}_i) = E(\bm{X}_i\mid \bm{Z}_i, \bm{W}_i)$. Then, in the literature, two related methods \citep{carrollMeasurementErrorNonlinear2006, rosnerCorrectionLogisticRegression1990}, both referred to as RC, use equation $E(\bm{x}_i\mid \bm{Z}_i, \bm{W}_i)$ to correct for measurement error. It has been shown that these two versions are algebraically equivalent for generalized linear models \citep{Thurston2003}. In this paper, we adopt the latter for discussion. 

Following \citet{rosnerCorrectionLogisticRegression1989}, the estimated parameter $\widehat{\bm{\Gamma}}_1$ from model \eqref{eq:MEM} is used to adjust the naive estimator $\widehat{\bm{\beta}}_1^*$. The rationale builds on the surrogacy assumption \citep{carrollMeasurementErrorNonlinear2006}, stating that measurement error is conditionally independent of the outcome given the true exposures. For linear regression, the Cox model \citep{spiegelmanRegressionCalibrationMethod1997}, logistic regression \citep{rosnerCorrectionLogisticRegression1990, rosnerCorrectionLogisticRegression1992}, and generalized linear models under small measurement error \citep{Raymond1990}, this correction can be expressed as
\begin{equation}
\widehat{\bm{\beta}}_1 = \widehat{\bm{\Gamma}}_1^{-1} \widehat{\bm{\beta}}_1^*.
\label{eq:RC}
\end{equation}
Since $\widehat{\bm{\beta}}_1^*$ is estimated directly from the main study, no transportability issue arises. To ensure the validity of the RC method, the transportability condition is always directly related to $\widehat{\bm{\Gamma}}_1$. Specifically, under the assumption that the measurement error generating process is similar across studies, existing literature typically requires that the conditional distribution of $\bm{x}$ given $\bm{W}$ be identical across the main study and the EVS \citep{Wong2020,li2025}. However, as demonstrated in Sections 2.3 and 3, this is only a sufficient condition. Applying this criterion, some scenarios that actually satisfy transportability may be incorrectly classified as violations. In this work, we aim to refine this understanding by identifying which components of $\bm{x}$ influence the transportability. 

\subsection{Measurement Error Mechanism}

To answer this question, we consider 
\begin{align}
    \bm{x}_i &= \bm{a}_{0,m} + \bm{A}_{2,m}^\T \bm{W}_i + \bm{\epsilon}_{m}, i=1,\ldots, n_m,\label{eq:xm}\\
    \bm{x}_i &= \bm{a}_{0,v} + \bm{A}_{2,v}^\T \bm{W}_i + \bm{\epsilon}_{v}, i=n_m+1,\ldots, n_m+n_v,\label{eq:xv}
\end{align}
for the main study and the EVS, respectively. Here, $\bm{a}_{0,m}$ and $\bm{a}_{0,v}$ are $p$-dimensional vectors, $\bm{A}_{2,m}$ and $\bm{A}_{2,v}$ are $q\times p$ matrices, $\bm{\epsilon}_{m}$ has mean 0 and variance $\bm{\Sigma}_{m}$, and $\bm{\epsilon}_{v}$ has mean 0 and variance $\bm{\Sigma}_{v}$. Since $\bm{x}$ is correlated with confounders, this specification enables a finer decomposition of $\bm{x}$ to determine which component affects the transportability. 

Building on models \eqref{eq:xm} and \eqref{eq:xv}, explicitly defining the measurement error generating process is essential to link all variables and enable the derivation of a closed-form expression for $\bm{\Gamma}_1$. Following the framework proposed by \cite{Spiegelman1997}, we consider
\begin{align}
    \bm{X}_i &= \bm{x}_i + \bm{\epsilon}_{b},\label{eq:MEMB}\\
    \bm{Z}_i &= \bm{c}_0 + \bm{C}_1^\T \bm{x}_i + \bm{C}_2^\T \bm{W}_i + \bm{\epsilon},\label{eq:MEMC}
\end{align}
where $\bm{c}_0$ is a $p$-dimensional vector, $\bm{C}_1$ and $\bm{C}_2$ are $p\times p$ and $q\times p$ matrices, respectively. The error terms $\bm{\epsilon}_{b}$ and $\bm{\epsilon}$ are mean zero with variances $(\bm{\Sigma}_{b})_{p\times p}$ and $(\bm{\Sigma})_{p\times p}$, respectively. Notably, when $\bm{\Sigma}_{b} = \bm{0}$, the model simplifies to a classical-like error model, specifying the conditional distribution of $\bm{Z}$ given $(\bm{X}, \bm{W})$. On the other hand, when $\bm{\Sigma} = \bm{0}$, it reduces to a Berkson-like error model, which corresponds exactly to model \eqref{eq:MEM}. If measurement error affects variables as in model \eqref{eq:MEM}, no transportability issue arises. Therefore, in the rest of this paper, we assume $\bm{\Sigma} \ne \bm{0}$. 

\subsection{Validity Condition of RC}

Using the parameters from models \eqref{eq:xm}-\eqref{eq:MEMC}, the RC estimator for $\bm{\beta}_1$ can be shown to converge in probability to
\begin{equation}
    (\bm{C}_1 \bm{\Sigma}^{-1} \bm{C}_1^T + \bm{\Sigma}_v^{-1})(\bm{C}_1 \bm{\Sigma}^{-1} \bm{C}_1^T + \bm{\Sigma}_m^{-1})^{-1}\bm{\beta}_1.\label{PE_RC}
\end{equation}
The detailed derivation is provided in the Supplementary Material A.1. Here, we outline the key step. The argument relies on a semi-parametric approach that applies two linear transformation matrices to obtain the closed-form solution, rather than assuming normality on the variables. This strategy ensures that the resulting expression remains valid regardless of the underlying variable distributions. 

When $\bm{\Sigma}_v = \bm{\Sigma}_m$, this closed form in \eqref{PE_RC} indicates the RC estimator converges to $\bm{\beta}_1$ without bias. In other words, the validity of the RC estimator is determined by the conditional variance of $\bm{x}$ given $\bm{W}$, rather than by other factors such as the distributional shape or mean of $\bm{x}$, or the distribution of $\bm{W}$. For example, when $\bm{x}$ given $\bm{W}$ follows a gamma distribution in the EVS but a normal distribution in the main study, if the conditional variance of $\bm{x}$ given $\bm{W}$ remains the same across studies, RC can still provide valid estimates, as illustrated in Section 3. Furthermore, since the necessary condition is characterized by the conditional variance of $\bm{x}$ given $\bm{W}$, the conditional distribution of $\bm{Z}$ given $\bm{W}$, the indicator of the transportability issue proposed by \cite{li2025}, can be simplified to the conditional variance of $\bm{Z}$ given $\bm{W}$. 

In addition, when $\bm{x}_i$, $\bm{X}_i$, and $\bm{Z}_i$ are assumed to be scalars, the relative biases of the RC and naive estimators can be derived (See details in Supplementary Material A.2). Specifically, the models simplify as follows. In the main study, $x_i \sim (a_{0,m} + A_{2,m}W_i, \sigma_m^2)$ and $Z_i \sim (c_0 + C_1x_i + C_2W_i, \sigma^2)$. In the EVS, $x_i \sim (a_{0,v} + A_{2,v}W_i, \sigma_v^2)$ and $Z_i \sim (c_0 + C_1x_i + C_2W_i, \sigma^2)$. Here, the notation $(\mu, \sigma^2) $ specifies a distribution with mean $\mu$ and variance $\sigma^2$. This setting allows us to examine the conditions under which the naive method yields smaller bias than the RC estimator, as summarized in Table \ref{condi}. In general, the RC method usually but not always exhibits smaller bias than the naive method. In extreme scenarios, such as when there is a significant difference between the main study and the EVS, the naive estimator may yield a smaller bias, though such situations are likely rare in practice. 

\begin{table}[htbp]
\caption{Conditions for scalar variable under different situations. In the main study, $x_i\sim(a_{0,m}+A_{2,m}W_i,\sigma_m^2), Z_i\sim (c_0 + C_1x_i + C_2W_i,\sigma^2)$. In the external validation study, $x_i\sim(a_{0,v}+A_{2,v}W_i,\sigma_v^2), Z_i\sim (c_0 + C_1x_i + C_2W_i,\sigma^2)$. Here, the notation $(\mu, \sigma^2) $ specifies a distribution with mean $\mu$ and variance $\sigma^2$.}
\label{condi}
\begin{center}
\begin{tabular}{|p{0.95\textwidth}|}
\hline
Conditions under which the naive method yields a smaller bias\\
\hline
1. $C_1 = 0.5 \text{ and } \sigma^2 = \sigma_{m}^2/4$. \\
2. $C_1 = \frac{1 \pm \sqrt{1 - 4\sigma^2/\sigma_{m}^2}}{2}  \text{ and } \sigma^2 < \sigma_{m}^2/4$. \\
3. $\sigma^2 > \sigma_{m}^2/4\text{ and } \sigma_{v}^2 < \min \{\sigma_{m}^2, \frac{\sigma^2\sigma_{m}^2}{2\sigma^2 + (C_1^2 - C_1)\sigma_{m}^2}\}$. \\
4. $C_1 \in (-\infty,\frac{1 - \sqrt{1 - 4\sigma^2/\sigma_{m}^2}}{2})\cup (\frac{1 + \sqrt{1 - 4\sigma^2/\sigma_{m}^2}}{2},\infty), \sigma^2 < \sigma_{m}^2/4,$\\
and $\sigma_{v}^2 < \min \{\sigma_{m}^2, \frac{\sigma^2\sigma_{m}^2}{2\sigma^2 + (C_1^2 - C_1)\sigma_{m}^2}\}$. \\
5. $C_1 \in (\frac{1 - \sqrt{1 - 4\sigma^2/\sigma_{m}^2}}{2}, \frac{1 + \sqrt{1 - 4\sigma^2/\sigma_{m}^2}}{2}), \sigma^2 < \sigma_{m}^2/4,$\\
and $\sigma_{v}^2 < \min \{\sigma_{m}^2, \frac{\sigma^2}{C_1 - C_1^2}\}$. \\
6. $C_1 \in (\frac{1 - \sqrt{1 - 4\sigma^2/\sigma_{m}^2}}{2}, \frac{1 + \sqrt{1 - 4\sigma^2/\sigma_{m}^2}}{2}), \frac{C_1 - C_1^2}{2}\sigma_{m}^2 < \sigma^2 < \sigma_{m}^2/4,$\\
and $\sigma_{v}^2 > \max \{\sigma_{m}^2, \frac{\sigma^2\sigma_{m}^2}{2\sigma^2 + (C_1^2 - C_1)\sigma_{m}^2}\}$.\\
\hline
Conditions for $\vert\text{Relative Bias}_{\text{RC}}\vert \le R$\\
\hline
1. $\frac{\sigma^2 \sigma_{m}^2}{\sigma^2 + RC_1^2\sigma_{m}^2 + R\sigma^2}\le \sigma_{v}^2 < \sigma_{m}^2$. \\
2. $\sigma_{m}^2 < \sigma_{v}^2 \le \frac{\sigma^2 \sigma_{m}^2}{\sigma^2 + RC_1^2\sigma_{m}^2 + R\sigma^2}, \text{ if }R\le \frac{\sigma^2}{C_1^2\sigma_{m}^2 + \sigma^2}$. \\
\hline

\end{tabular}
\end{center}
\end{table}

\begin{figure}
    \centering
    \subfigure[Setting: $C_1 = 0.5, \sigma^2 = 0.25$, and $\sigma_m^2 = 1$]{
    \includegraphics[width=0.45\linewidth]{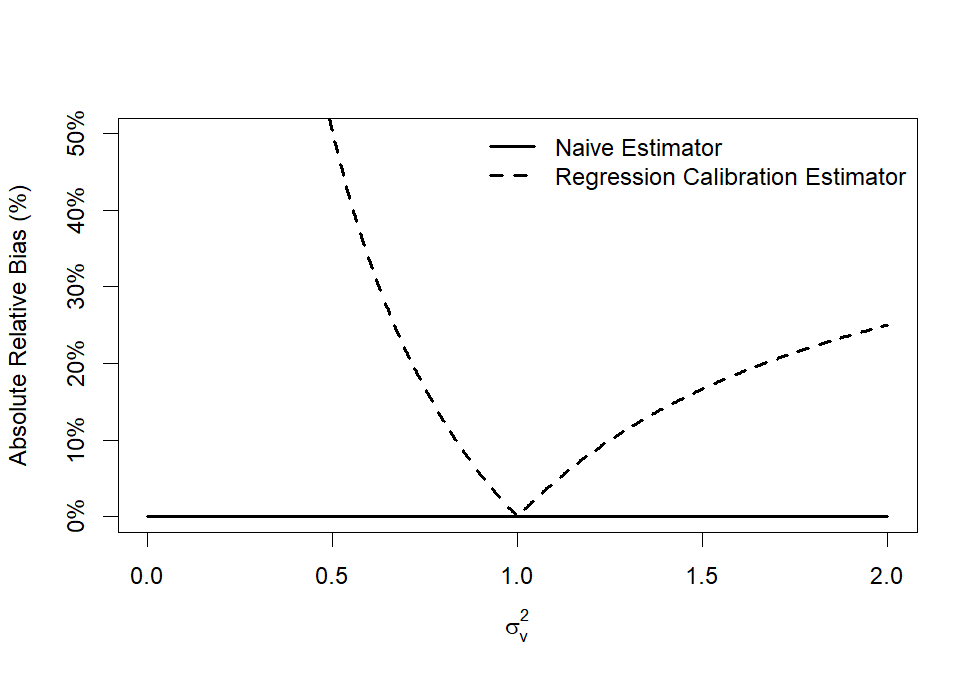}}
    \subfigure[Setting: $C_1 = 1, \sigma^2 = 0.4$, and $\sigma_m^2 = 1$]{
    \includegraphics[width=0.45\linewidth]{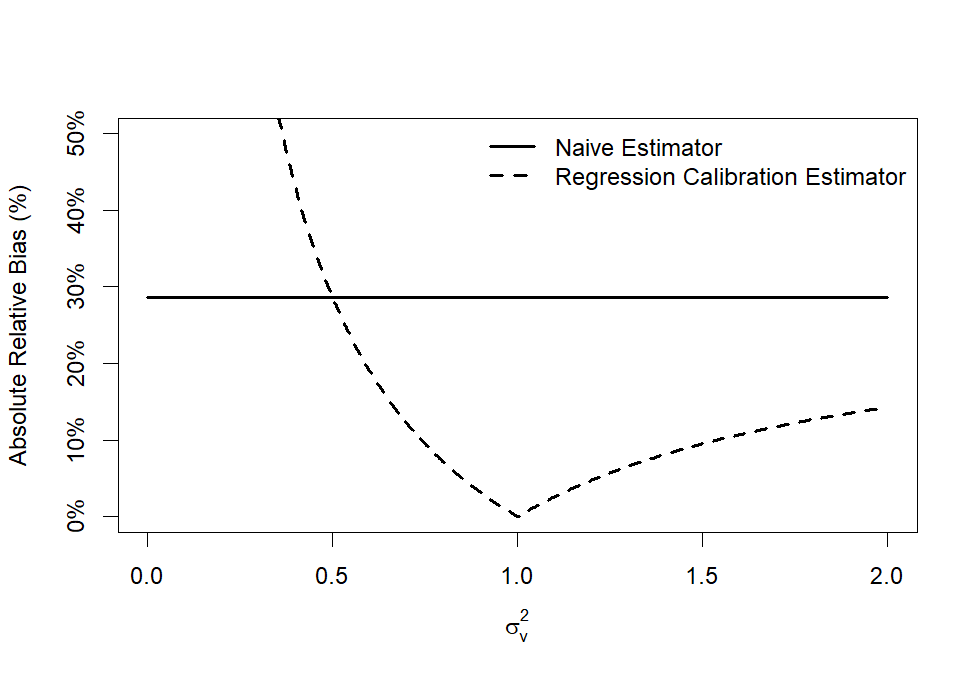}}
    \subfigure[Setting: $C_1 = 0.5, \sigma^2 = 0.2$, and $\sigma_m^2 = 1$]{
    \includegraphics[width=0.45\linewidth]{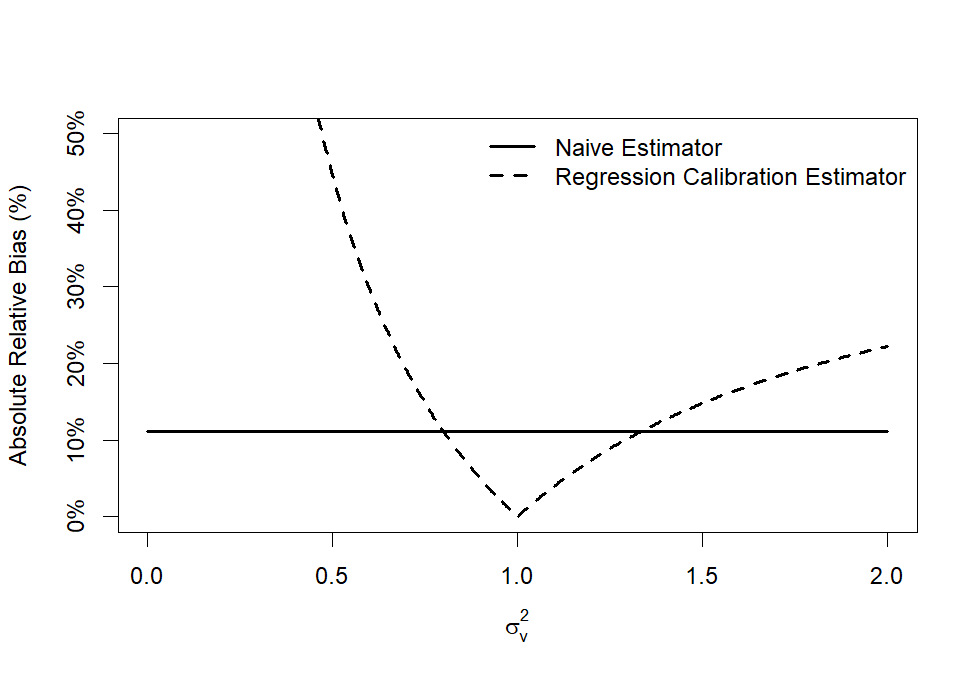}}
    \subfigure[Setting: $C_1 = 1, \sigma^2 = 0.1$, and $\sigma_m^2 = 1$]{
    \includegraphics[width=0.45\linewidth]{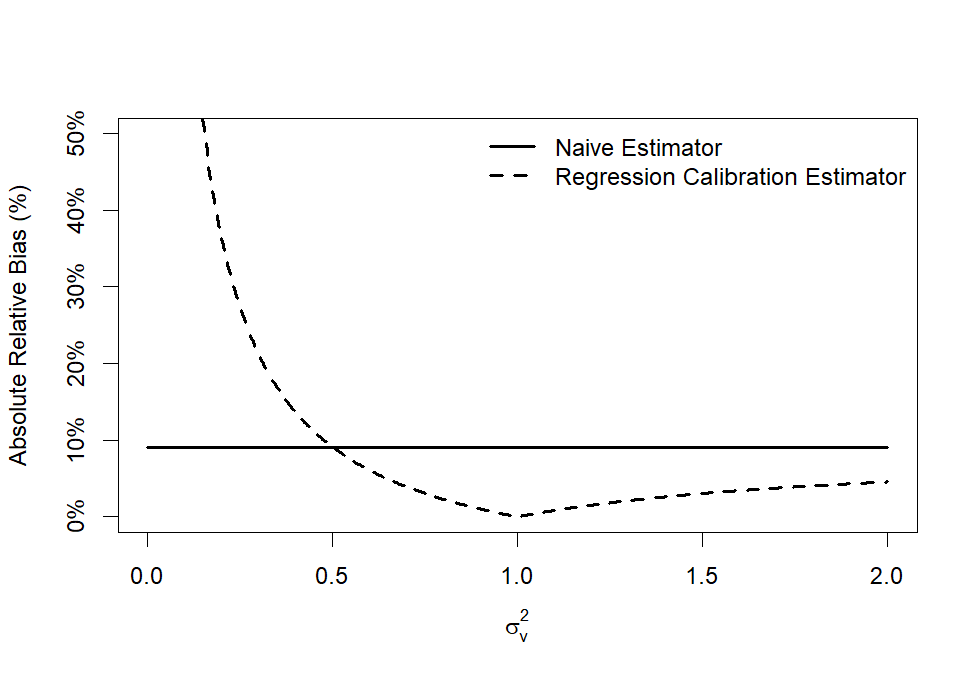}}
    \caption{Graphs showing the relative bias of the naive estimator (solid), and the relative bias of the RC method (dashes). In the main study, $x_i\sim(a_{0,m}+A_{2,m}W_i,\sigma_m^2), Z_i\sim (c_0 + C_1x_i + C_2W_i,\sigma^2)$. In the external validation study, $x_i\sim(a_{0,v}+A_{2,v}W_i,\sigma_v^2), Z_i\sim (c_0 + C_1x_i + C_2W_i,\sigma^2)$.}
    \label{Fig:RB}
\end{figure}

Moreover, four representative patterns of relative bias for the RC and naive approaches under different parameter settings were illustrated in Figure \ref{Fig:RB}. When $C_1 = 0.5$ and $\sigma^2 = \sigma_{m}^2/4$, the relative bias of the naive estimator equals zero. Under this setting, the naive estimator has a smaller bias than the RC estimator. However, this represents a highly special case that occurs with probability zero. Except for this special case and other extreme scenarios, under modest departures from the transportability, the RC estimator consistently exhibits lower bias than the naive estimator.

\section{Simulation Studies}

We designed simulation studies for both continuous (Scenario 1) and binary (Scenario 2) outcomes to illustrate two key points in Section 2: (1) the validity of the RC method depends solely on the conditional variance of the true exposure given confounders, and (2) the RC estimator usually exhibits smaller bias than the naive approach when the transportability assumption does not hold, especially under modest departures from this assumption. The simulation setup used $n_m = 10,000$ to reflect the scale of epidemiologic cohorts and $n_v = 500$, similar to the Men’s Lifestyle Validation Studies \citep{MLVS}. Here, smaller sample sizes for the main study were not considered because modern epidemiologic cohorts applying the RC method usually include more than $10,000$ subjects \citep{samplesize1, samplesize2}. 

The data generation procedure was structured as follows. Throughout, we assumed an error-free continuous confounder $W$ generated from a normal distribution $\mathcal{N}(1,1)$. Scalar true exposures $x$ in the main study and the EVS were obtained using models \eqref{eq:xm} and \eqref{eq:xv}, respectively. Then, model \eqref{eq:MEMB} was applied to generate the scalar alloyed gold standard exposure $X$ in both studies, and model \eqref{eq:MEMC} was used to create the scalar surrogate exposure $Z$ in both studies. In Scenarios 1 and 2, the outcome $Y$ was produced from the following linear and logit models, respectively. Specifically,
\begin{align*}
    & Y_i \mid x_i, W_i \sim \mathcal{N}(\beta_0 + \beta_1 x_i + \beta_2 W_i, 1), i=1,\ldots,n_m,\\
    & Y_i \mid x_i, W_i \sim \text{Bern}(p_i), \text{ where }\log\{p_i/(1-p_i)\} = \beta_0 + \beta_1 x_i + \beta_2 W_i, i=1,\ldots,n_m.
\end{align*}
In all simulations, we set $\beta_0 = -5, \beta_1 = 1.0, \beta_2 = 0.5, c_0 = 1, C_1 = 1, C_2 = 0.5, A_{2,m} = 0.1, a_{0,m} = a_{0,v}=0$, and assumed that all variables in the main study followed normal distributions. In addition, the standard deviation of $x$ given $W$ in the main study is $\sigma_m = 1$, the standard deviation of the Berkson-like error is $\sigma_b = 0.1$. The standard deviation of the classical-like measurement error was set to $\sigma = 1.2$ in the continuous outcome scenario and $\sigma = 0.6$ in the binary outcome scenario. Under these settings, the condition $\beta_1^2 \widehat{\hbox{Var}}(x \mid Z, W) < 0.5$  \citep{Kuha1994} holds in the binary scenario, indicating a low level of measurement error and ensuring the validity of equation \eqref{eq:RC}. This design allows us to examine the impact of the transportability assumption without introducing other sources of bias. On the other hand, since the continuous scenario does not require a small measurement error assumption, we considered a setting in which $\beta_1^2 \widehat{\hbox{Var}}(x \mid Z, W) > 0.5$.  

In each scenario, we investigated four cases that varied in the settings used to generate $x$ in the EVS. Specifically, in the first case, all parameter values were identical to those in the main study, and all variables also followed normal distributions. In contrast, in the second case, both the conditional mean and the distributional shape of $x$ given $W$ were changed in the EVS, where $A_{2,v} = 1.2 A_{2,m}$ and $x\mid W$ was drawn from a gamma distribution with shape parameter $\sigma_m^2$ and rate parameter 1, securing $\hbox{Var}(x\mid W)$ the same across both studies but the shape in the EVS is significantly far from normal, and then centered by subtracting $\sigma_m^2$. In the other two cases, all other settings were similar in the first case, while we assigned different values to $\sigma_v$, i.e., $\sigma_v^2 = 1.2 \sigma_m^2$ and $\sigma_v^2 = 1.5 \sigma_m^2$, representing increasing degrees of violation of the transportability assumption. All simulations were repeated 10,000 times. 

The results for two scenarios were reported in Table \ref{simu}, including the average point estimates and relative biases. As expected, the RC method consistently exhibited low bias in the first two cases. This pattern verified the theoretical result that, although the distribution of $x$ given $W$ is different between the main study, as long as the $\hbox{Var}(x\mid W)$ is the same, the RC estimator is still valid. On the other hand, in cases 3 and 4, the relative bias of the RC method increased as the setting deviated further from the transportability assumption, i.e., the $\hbox{Var}(x\mid W)$ differed more substantially from that in the main study. This trend aligns with Figure \ref{Fig:RB} in Section 2. In addition, in all scenarios, the RC estimator consistently showed lower bias than the naive method, further supporting the conclusion derived in Section 2. 

\begin{table}[htbp]
  \centering
  \caption{Simulation results for both continuous (Scenario 1) and binary (Scenario 2) outcomes. TV is the true value of $\beta_1$, and PE is the average of the point estimates for $\beta_1$ across simulations. Bias (\%) represents the relative bias, calculated as $(\text{PE} - \text{TV}) / \text{TV} \times 100$. RC denotes the regression calibration approach. All settings in the external validation study are identical to the main study in Case 1. Case 2 changed both the conditional mean and the distributional shape of true exposure given confounders in the external validation study. While Cases 3 and 4 only changed the conditional variance of true exposure given confounders in the external validation study, i.e., $\sigma_v^2 = 1.2 \sigma_m^2$ for Case 3 and $\sigma_v^2 = 1.5 \sigma_m^2$ for Case 4. }
    \begin{tabular}{rcccc}
    \hline
    \multicolumn{4}{l}{Scenario 1: Continuous outcomes} \\
    \hline
    & TV & Method & PE & Bias(\%) \\
    \hline
    Case 1 & 1.0   & RC    & 1.00  & 0.4\% \\
          &       & Naive & 0.41  & -59.0\% \\
    Case 2 & 1.0   & RC    & 1.01  & 1.0\% \\
          &       & Naive & 0.41  & -59.0\% \\
    Case 3 & 1.0   & RC    & 0.90  & -9.6\% \\
          &       & Naive & 0.41  & -59.0\% \\
    Case 4 & 1.0   & RC    & 0.80  & -19.5\% \\
          &       & Naive & 0.41  & -59.0\% \\
    \hline
    \multicolumn{4}{l}{Scenario 2: Binary outcomes} \\
    \hline
    & TV & Method & PE & Bias(\%) \\
    \hline
    Case 1 & 1.0   & RC    & 0.99  & -1.1\% \\
          &       & Naive & 0.73  & -27.3\% \\
    Case 2 & 1.0   & RC    & 0.99  & -0.8\% \\
          &       & Naive & 0.73  & -27.3\% \\
    Case 3 & 1.0   & RC    & 0.94  & -5.5\% \\
          &       & Naive & 0.73  & -27.4\% \\
    Case 4 & 1.0   & RC    & 0.90  & -9.7\% \\
          &       & Naive & 0.73  & -27.2\% \\
    \hline
    \end{tabular}%
  \label{simu}%
\end{table}%

\section{Illustrative Example}

We applied both the RC and naive estimators to investigate the effect of moderate physical activity on the risk of cardiovascular disease (CVD) using data from the Health Professionals Follow-Up Study (HPFS) \citep{HPFS1999}. The HPFS is a prospective cohort study that includes 51,529 male health professionals aged 40 to 75 years enrolled in 1986. In our analysis, the HPFS served as the main study, where the outcome, the surrogate exposure, and relevant confounders were available. Moderate physical activity (MET hours per day), assessed by the physical activity questionnaire (PAQ) in 2010, was used as the baseline surrogate exposure $Z$. We selected 2010 because the PAQ from that year was the earliest version identical in structure to that used in the validation data, ensuring consistency in questionnaire design and minimizing potential differences in the measurement error generating process across studies. After excluding men with a history of CVD or missing data on moderate physical activity in 2010, a total of 18,510 subjects were included in the analysis. Among them, 550 developed CVD between 2010 and 2016. 

In practice, confounders must be carefully selected when conducting causal analyses using the RC method. \cite{Tang2024} demonstrated that confounders should be considered in models \eqref{eq:outcome2} and \eqref{eq:MEM} to avoid bias in the RC estimator. Therefore, we only included confounders, $\bm{W}$, such as aspirin use, intake of polyunsaturated fat, omega-3 fatty acids, alcohol, and fiber. These variables are a subset of baseline covariates considered by \cite{Chomistek2012} and are available in both the main study and the EVS. 

To perform the RC method, the surrogate exposure was validated using physical activity diaries, which is the reference measurement $X$. In the Men’s Lifestyle Validation Study (MLVS) \citep{MLVS}, moderate physical activity was assessed by both the PAQ and the more accurate diaries. The MLVS included 671 men from the HPFS and Harvard Pilgrim Health Care (HPHC), conducted between 2011 and 2013 (484 from HPFS and 187 from HPHC). This design, comprising two sub-validation studies, provides a useful tool for illustrating conclusions related to transportability. Specifically, subjects in the HPFS validation sample were randomly selected from individuals who had completed the main study's questionnaire in 2006/2007 and had given blood samples in 1994. It is therefore reasonable to expect that the transportability problem between this validation study and the main study would be minimal. On the other hand, subjects in the HPHC were randomly selected from another group of men, which may lead to potential transportability concerns related to the main study. In our analysis, we used these two datasets separately to apply the RC method and to illustrate our findings. 

The variable characteristics and final analysis results were reported in Table \ref{real}. First of all, the difference in the conditional variance of $Z$ given $W$ between the HPFS and HPHC validation studies was comparable to the difference in the conditional variance of $X$ given $W$ between these two studies, suggesting that the conditional variance of $Z$ given $W$ serves as a reasonable indicator of the transportability issue. In addition, as expected, the transportability issue does not exist between the HPFS validation data and the main study since the value of the conditional variance of $Z$ given $W$ in the HPFS validation study closely matched that observed in the main study. In contrast, the conditional variance of $Z$ given $W$ in the HPHC was lower than the corresponding value in the main study. Based on this observation, the RC estimator obtained using the HPFS validation study can be regarded as the value close to the true value. Then, the RC estimator derived from the HPHC still exhibits a smaller bias than the naive estimator when the transportability assumption is violated. 

\begin{table}[htbp]
  \centering
  \caption{Variable characteristics and analysis results. The main study is the Health Professionals Follow-Up Study, HPFS-VS is the validation data of the Health Professionals Follow-Up Study, and HPHC-VS is the validation data from the Harvard Pilgrim Health Care. 95\% CI is the 95\% confidence interval. RC-HPFS denotes the regression calibration estimator using HPFS-VS, and RC-HPHC refers to the regression calibration estimator obtained from HPHC-VS. }
    \begin{tabular}{lcccc}
    \hline
    \multicolumn{5}{l}{Variable Characteristics} \\
    \hline
          & Mean($Z$) & Var($Z\mid \bm{W}$) & Mean($X$) & Var($X\mid\bm{W}$) \\
    \hline
    Main Study & 2.77 & 4.56 &       &  \\
    HPFS-VS & 5.06 & 4.26 & 7.78 & 5.75 \\
    HPHC-VS & 3.82 & 3.35 & 6.67 & 4.88 \\
    \hline
    \multicolumn{5}{l}{Analysis Results} \\
    \hline
          & Estimate & Relative Bias (\%) &  95\% CI     &  \\
    \hline
    RC - HPFS & -0.19 &       &    (-0.31, -0.07)   &  \\
    RC - HPHC & -0.10 & -45.5\% &  (-0.17, -0.04)     &  \\
    Naive & -0.06 & -68.9\% &  ( -0.09, -0.03 )     &  \\
    \hline
    \end{tabular}%
  \label{real}%
\end{table}%

\section{Discussion}

In this work, we revisited the RC method, a widely used approach for correcting measurement error \citep{Shaw2018}. A critical step in applying RC is to determine whether the transportability assumption holds, that is, whether the MEM estimated from the EVS can be validly transferred to the main study. In current practice, assuming the distributions of $\bm{X}$ given $\bm{W}$ are the same across studies is commonly used \citep{Wong2020}, in which case the conditional distribution of $\bm{Z}$ given $\bm{W}$ has been proposed as an indicator for evaluating it \citep{li2025}. However, this criterion may incorrectly classify some transportable scenarios, especially those involving differences in the shape or mean of the distribution of $\bm{X}$, as violations. To address this limitation, we derived the sufficient and necessary condition for transportability: the variance of $\bm{X}$ given $\bm{W}$ should be the same across studies. Building on this, the indicator can be simplified to the conditional variance of $\bm{Z}$ given $\bm{W}$. Finally, we explored the conditions under which the RC estimator exhibits less bias than the naive estimator when the exposure of interest is scalar. It turns out that the RC method is usually more reliable than the naive method in most practical settings. 

Although the transportability issue of the analysis in the main study is typically mentioned in the context of the EVS, it can also arise in internal validation study designs. Specifically, when the internal validation study is not selected completely at random from the main study, as is often the practice case \citep{NHSV1985, HPFSV1992}, the validity of the RC estimator may also be in question. \citet{Spiegelman2001} derived the conditions under which RC remains valid when the sampling mechanism for the internal validation study is not completely random. In situations where these conditions are not satisfied, the indicator, variance of $\bm{Z}$ given $\bm{W}$, should also be applied to assess the transportability. 

In addition, we acknowledge that our review of the RC method in Section 2.1 does not cover all research contexts, for instance, cases where the error terms $\bm{\epsilon}_b$ in model \eqref{eq:MEMB} and $\bm{\epsilon}$ in model \eqref{eq:MEMC} are correlated. Our focus was restricted to the setting where these errors are independent, primarily because this assumption simplifies the connection between $\bm{X}\mid \bm{Z, W}$ in the EVS and the working model $\bm{x} \mid \bm{Z}, \bm{W}$ used in the derivation. In practice, however, our results are not limited to this simplified scenario. Regardless of the correlation structure among errors, valid estimates of $\bm{x}\mid \bm{Z, W}$ remain feasible, and our results continue to hold for various versions of the RC method built upon this model. For example, when $\bm{x} \mid \bm{Z}$ is estimable by including additional biomarkers or alternative measurements \citep{Spiegelman1997, Spiegelman2005}, the residual method \citep[Chapter 11]{willettNutritionalEpidemiology2012} can then be used to incorporate $\bm{W}$. 

\bibliographystyle{plainnat} 
\bibliography{ref}  
 
\section*{A Theoretical derivations}

\subsection*{A.1 Closed form of $\widehat{\bm{\beta}}_1$ in Regression Calibration}

As in the main text, the models are
\begin{align}
    &g\{\mathbb{E}(Y_i|\bm{Z}_i,\bm{W}_i)\} = \beta_0^* + \bm{Z}_i^T \bm{\beta}_1^*  + \bm{W}_i^T\bm{\beta}_2^*,\label{outcome}\\
    &\bm{Z}_i=\bm{c}_0 + \bm{C}_1^\T \bm{x}_i + \bm{C}_2^\T \bm{W}_i + \bm{\epsilon},\label{MC}\\
    &\bm{x}_i = \bm{a}_{0,v} + \bm{A}_{2,v}^\T \bm{W}_i + \bm{\epsilon}_{v}, \label{Mx}\\
    &E(\bm{x}_i\mid \bm{Z}_i, \bm{W}_i) = \bm{\gamma}_{0} + \bm{\Gamma}_{1}^\T \bm{Z}_i + \bm{\Gamma}_{2}^\T \bm{W}_i\label{MEM}.
\end{align}

As discussed in the main text, we need to investigate how violations of the transportability assumption influence ${\bm{\Gamma}}_1$. Let ${\bm{\Gamma}}_{1,m}$ be the value in the main study, and ${\bm{\Gamma}}_{1,v}$ be the value in the external validation study. Then, for the RC estimator $\widehat{\bm{\beta}}_1$, we have
\begin{equation}
    \widehat{\bm{\beta}}_1 = \widehat{\bm{\Gamma}}_1^{-1} \widehat{\bm{\beta}}_1^* \rightarrow_p {\bm{\Gamma}}_{1,v}^{-1}{\bm{\Gamma}}_{1,m}{\bm{\beta}}_1.\label{RChat}
\end{equation}
If ${\bm{\Gamma}}_{1,v}$ and ${\bm{\Gamma}}_{1,m}$ can be expressed in terms of the parameters from models \eqref{MC} and \eqref{Mx}, the closed-form expression of $\widehat{\bm{\beta}}_1$ can be derived.

The general idea is to establish a connection between model \eqref{MEM} and models \eqref{MC} and \eqref{Mx}, allowing $\bm{\Gamma}_1$ in model \eqref{MEM} to be expressed in terms of the parameters from models \eqref{MC} and \eqref{Mx}. Specifically, for the external validation study, we can find linear transforms $\bm{L}_1, \bm{L}_2$ to combine model \eqref{MC} and \eqref{Mx} which is:
\begin{equation}
    \bm{L}_1( \bm{c}_0 + \bm{C}_1^\T \bm{x}_i + \bm{C}_2^\T \bm{W}_i + \bm{\epsilon}) + \bm{L}_2 \bm{x}_i = \bm{L}_1\bm{Z}_i + \bm{L}_2(\bm{a}_{0,v} + \bm{A}_{2,v}^\T \bm{W}_i + \bm{\epsilon}_{v}). \label{eq: 1}
\end{equation}
The distribution of $\bm{x}_i$ given $\bm{Z}_i$ and $\bm{W}_i$ in the external validation study can then be written as
\begin{equation}
		\bm{x}_i = (\bm{L}_1\bm{C}_1^T + \bm{L}_2)^{-1}  \{(\bm{L}_2\bm{a}_{0,v} - \bm{L}_1\bm{c}_0 ) + \bm{L}_1\bm{Z}_i + (\bm{L}_2\bm{A}_{2,v}^T - \bm{L}_1\bm{C}_2^T)\bm{W}_i + (\bm{L}_2 \bm{\epsilon}_{v} - \bm{L}_1\bm{\epsilon})\}. \label{eq: 2}
\end{equation}
Note that the relationship between $\bm{x}_i$ and $(\bm{Z}_i, \bm{W}_i)$ in \eqref{eq: 2} holds for any choice of linear operators $\bm{L}_1$ and $\bm{L}_2$. Conceptually, model \eqref{MEM} is a special member of the class of linear models defined by \eqref{eq: 2} and is expected to explain the greatest proportion of variation in $\bm{x}_i$ among them. Therefore, to obtain the parameters in \eqref{MEM}, we identify the linear model within \eqref{eq: 2} that minimizes the error variance, i.e., for any $p$-dimensional vector $\bm{\alpha}$,
\begin{equation*}
    \min_{\bm{L}_1,\bm{L}_2} \bm{\alpha}^T(\bm{L}_1\bm{C}_1^T + \bm{L}_2)^{-1} (\bm{L}_2\bm{\Sigma}_v\bm{L}_2^T + \bm{L}_1\bm{\Sigma}\bm{L}_1^T)((\bm{L}_1\bm{C}_1^T + \bm{L}_2)^{-1} )^T\bm{\alpha}.
\end{equation*}
The solution is 
$$\bm{L}_1 = k\bm{C}_1 \bm{\Sigma}^{-1}, \quad \bm{L}_2 = k\bm{\Sigma}_v^{-1}, \quad k \in \mathbb{R}. $$
	\begin{proof}
		Let's define
		\begin{equation*}
			f = \bm{ABA}^T = \operatorname{trace}(\bm{ABA}^T), \text{ where }\bm{A} = \bm{\alpha}^T(\bm{L}_1\bm{C}_1^T + \bm{L}_2)^{-1}, \bm{B} = (\bm{L}_2\bm{\Sigma}_v\bm{L}_2^T + \bm{L}_1\bm{\Sigma}\bm{L}_1^T)
		\end{equation*}
		
		Then we have
		\begin{equation}
			\begin{aligned}
				df &= \operatorname{trace}((d\bm{A})\bm{BA}^T) + \operatorname{trace}(\bm{A}(d\bm{B})\bm{A}^T) + \operatorname{trace}(\bm{AB}(d\bm{A})^T)\\
				&= \operatorname{trace}((\bm{B}+\bm{B})^T\bm{A}^Td\bm{A}) + \operatorname{trace}(\bm{A}^T\bm{A}d\bm{B}) \label{a1}
			\end{aligned}
		\end{equation}
		
		Based on the equation (\ref{a1}) and the definition of differential which is
		\begin{equation}
			df = \operatorname{trace}\left(\frac{\partial f^T}{\partial \bm{A}}d\bm{A}\right) + \operatorname{trace}\left(\frac{\partial f^T}{\partial \bm{B}}d\bm{B}\right)\label{a2}
		\end{equation}
		we have
		\begin{equation}
			\frac{\partial f^T}{\partial\bm{A}} = (\bm{B}+\bm{B})^T\bm{A}^T,\quad \frac{\partial f^T}{\partial\bm{B}} = \bm{A}^T\bm{A}\label{a3}
		\end{equation}
		
		Plugging $\bm{A} = \bm{\alpha}^T(\bm{L}_1\bm{C}_1^T + \bm{L}_2)^{-1}$ and $\bm{B} = (\bm{L}_2\bm{\Sigma}_v\bm{L}_2^T + \bm{L}_1\Sigma\bm{L}_1^T)$ into equation (\ref{a2}):
		\begin{equation*}
			\begin{aligned}
				df 
				&= \operatorname{trace}\left(\frac{\partial f^T}{\partial \bm{A}}d\bm{A}\right) + \operatorname{trace}\left(\frac{\partial f^T}{\partial \bm{B}}d\bm{B}\right)\\
				&= \operatorname{trace}\left(\frac{\partial f^T}{\partial \bm{A}}d(\bm{\alpha}^T(\bm{L}_1\bm{C}_1^T + \bm{L}_2)^{-1})\right) + \operatorname{trace}\left(\frac{\partial f^T}{\partial \bm{B}}d(\bm{L}_2\bm{\Sigma}_v\bm{L}_2^T + \bm{L}_1\Sigma\bm{L}_1^T)\right)\\
				&= \operatorname{trace}((2\bm{\Sigma}\bm{L}_1^T\bm{A}^T\bm{A} - \bm{C}_1^T(\bm{\alpha}^T)^{-1}\bm{A}(\bm{B} + \bm{B}^T)\bm{A}^T\bm{A})d\bm{L}_1)\\
				&+ \operatorname{trace}((2\bm{\Sigma}_v\bm{L}_2^T\bm{A}^T\bm{A} - (\bm{\alpha}^T)^{-1}\bm{A}(\bm{B} + \bm{B}^T)\bm{A}^T\bm{A})d\bm{L}_2)
			\end{aligned}
		\end{equation*}
		then we have
		\begin{equation}
			\begin{aligned}
				\frac{\partial f^T}{\partial \bm{L}_1} &= 2\bm{\Sigma}\bm{L}_1^T\bm{A}^T\bm{A} - \bm{C}_1^T(\bm{\alpha}^T)^{-1}\bm{A}(\bm{B} + \bm{B}^T)\bm{A}^T\bm{A}\\
				\frac{\partial f^T}{\partial \bm{L}_2} &=
				2\bm{\Sigma}_v\bm{L}_2^T\bm{A}^T\bm{A} - (\bm{\alpha}^T)^{-1}\bm{A}(\bm{B} + \bm{B}^T)\bm{A}^T\bm{A}
			\end{aligned}
		\end{equation}
		Based on the Supremum and Infimum principle, $f$ has a minimum value, after set 
		$$\frac{\partial f^T}{\partial \bm{L}_1} = \frac{\partial f^T}{\partial \bm{L}_2} = \bm{0},$$
		we get the solution
		$$\bm{L}_1 = k\bm{C}_1 \bm{\Sigma}^{-1}, \quad \bm{L}_2 =k \bm{\Sigma}_v^{-1},\quad k\in \mathbb{R}. $$
	\end{proof}
    Plugging $\bm{L}_1$ and $\bm{L}_2$ into equation \eqref{eq: 2}, we have
    \begin{equation}
		\bm{\Gamma}_{1,v} = ((\bm{L}_1\bm{C}_1^T + \bm{L}_2)^{-1} \bm{L}_1)^T = \left( \bm{\Sigma}^{-1} \bm{C}_1^T\right)\left( \bm{C}_1 \bm{\Sigma}^{-1} \bm{C}_1^T + \bm{\Sigma}_v^{-1} \right)^{-1}. \label{Gamma}
	\end{equation}
    Similarly, we have
    \begin{equation}
		\bm{\Gamma}_{1,m} = \left( \bm{\Sigma}^{-1} \bm{C}_1^T\right)\left( \bm{C}_1 \bm{\Sigma}^{-1} \bm{C}_1^T + \bm{\Sigma}_m^{-1} \right)^{-1}. \label{Gm}
	\end{equation}

    Plugging equations \eqref{Gamma} and \eqref{Gm} into equation \eqref{RChat}, we have
    \begin{equation}
        \widehat{\bm{\beta}}_1 \rightarrow_p (\bm{C}_1 \bm{\Sigma}^{-1} \bm{C}_1^T + \bm{\Sigma}_v^{-1})(\bm{C}_1 \bm{\Sigma}^{-1} \bm{C}_1^T + \bm{\Sigma}_m^{-1})^{-1}\bm{\beta}_1.\label{RC}
    \end{equation}

\subsection*{A.2 Naive and Regression Calibration Estimators under the Scalar Scenario}

    Based on the expression provided in \eqref{Gm}, the scalar form of $\Gamma_1$ in the main study is
    \begin{equation*}
		{\Gamma}_1 = (C_1/\sigma^2)(C_1^2/\sigma^2 + 1/\sigma_{m}^2)^{-1},
	\end{equation*}
    which implies 
    \begin{equation*}
        \widehat{\beta}_1^* \rightarrow_p \Gamma_1 \beta_1 = (C_1/\sigma^2)(C_1^2/\sigma^2 + 1/\sigma_{m}^2)^{-1}\beta_1.
    \end{equation*}

    Similarly, based on \eqref{RC},
    \begin{equation*}
        \widehat{\beta}_1 = \widehat{\beta}_1^*/\widehat{\Gamma}_1 \rightarrow_p (C_1^2/\sigma^2 + 1/\sigma_{v}^2)(C_1^2/\sigma^2 + 1/\sigma_{m}^2)^{-1}\beta_1. 
    \end{equation*}
    
    Based on the above equations, relative bias was calculated separately for the naive method and the regression calibration (RC) method. The relative biases were
    \begin{equation*}
        \begin{aligned}
            & \text{Relative Bias}_{\text{Naive}} = \frac{\widehat{\beta}_1^* - \beta_1}{\beta_1} = \frac{C_1/\sigma^2}{C_1^2/\sigma^2 + 1/\sigma_{m}^2} - 1 = \frac{C_1/\sigma^2 - C_1^2/\sigma^2 - 1/\sigma_{m}^2}{C_1^2/\sigma^2 + 1/\sigma_{m}^2}\\
            & \text{Relative Bias}_{\text{RC}} = \frac{\widehat{\beta}_1 - \beta_1}{\beta_1} = \frac{C_1^2/\sigma^2 + 1/\sigma_{v}^2}{C_1^2/\sigma^2 + 1/\sigma_{m}^2} - 1 = \frac{1/\sigma_{v}^2 - 1/\sigma_{m}^2}{C_1^2/\sigma^2 + 1/\sigma_{m}^2}. 
        \end{aligned}
    \end{equation*}

    According to the above expression, the relative bias of both methods may take either positive or negative values, depending on the parameters $C_1, \sigma^2, \sigma_{m}^2$ and $\sigma_{v}^2$. To determine when the naive method provides a smaller bias estimate than the RC estimator, we compared the absolute values of their relative biases case by case. Specifically, 
    \begin{enumerate}
        \item When $\{C_1 = 0.5 \text{ and } \sigma^2 = \sigma_{m}^2/4\}$ or $\{C_1 = \frac{1 \pm \sqrt{1 - 4\sigma^2/\sigma_{m}^2}}{2}  \text{ and } \sigma^2 < \sigma_{m}^2/4\}$, the relative bias of the naive method equals to 0. Thus, in this scenario, the naive method yields a smaller bias than the RC estimator. 
        
        \item When $\{\sigma^2 > \sigma_{m}^2/4\}$ or $\{C_1 \in (-\infty,\frac{1 - \sqrt{1 - 4\sigma^2/\sigma_{m}^2}}{2})\cup (\frac{1 + \sqrt{1 - 4\sigma^2/\sigma_{m}^2}}{2},\infty) \text{ and } \sigma^2 < \sigma_{m}^2/4\}$, we have 
        \begin{equation*}
            (C_1 - C_1^2)/\sigma^2 - 1/\sigma_{m}^2 <0.
        \end{equation*}
        According to $C_1^2/\sigma^2 + 1/\sigma_{m}^2>0$, we can conclude that the relative bias of the naive method is always negative under this scenario. On the other hand, the relative bias of the RC estimator still may be either positive or negative, depending on the values of parameters $\sigma_{v}^2$ and $\sigma_{m}^2$. 

        \begin{enumerate}
        \item When $\sigma_{v}^2 < \sigma_{m}^2$, we have $1/\sigma_{v}^2 - 1/\sigma_{m}^2 > 0$, which implies that the relative bias of the RC method is positive. Under this situation, the RC estimator has a larger bias than the naive estimator if and only if
        \begin{equation*}
            \text{Relative Bias}_{\text{naive}} + \text{Relative Bias}_{\text{RC}} > 0.
        \end{equation*}
        Hence, finding the conditions under which 
        \begin{equation*}
            \text{Relative Bias}_{\text{naive}} + \text{Relative Bias}_{\text{RC}} > 0.
        \end{equation*}
        identifies scenarios where the naive method performs better than the RC method. 

        According to
        \begin{equation}\label{eq:pp}
            \begin{aligned}
                \text{Relative Bias}_{\text{naive}} + \text{Relative Bias}_{\text{RC}} &= \frac{C_1/\sigma^2 - C_1^2/\sigma^2 - 1/\sigma_{m}^2}{C_1^2/\sigma^2 + 1/\sigma_{m}^2} + \frac{1/\sigma_{v}^2 - 1/\sigma_{m}^2}{C_1^2/\sigma^2 + 1/\sigma_{m}^2}\\
                &= \frac{(C_1 - C_1^2)/\sigma^2 + 1/\sigma_{v}^2 - 2/\sigma_{m}^2}{C_1^2/\sigma^2 + 1/\sigma_{m}^2},
            \end{aligned}
        \end{equation}
        we have
        \begin{equation*}
            \text{Relative Bias}_{\text{naive}} + \text{Relative Bias}_{\text{RC}}>0 \Leftrightarrow (C_1 - C_1^2)/\sigma^2 + 1/\sigma_{v}^2 - 2/\sigma_{m}^2 > 0, 
        \end{equation*}
        which implies that
        \begin{equation*}
            \text{Relative Bias}_{\text{naive}} + \text{Relative Bias}_{\text{RC}}>0 \Leftrightarrow \sigma_{v}^2 < \frac{\sigma^2\sigma_{m}^2}{2\sigma^2 + (C_1^2 - C_1)\sigma_{m}^2}.  
        \end{equation*}

        Therefore, when
        \begin{equation*}
            \sigma_{v}^2 < \min \{\sigma_{m}^2, \frac{\sigma^2\sigma_{m}^2}{2\sigma^2 + (C_1^2 - C_1)\sigma_{m}^2}\}, 
        \end{equation*}
        the naive method results in less bias than the RC estimator. 

        \item When $\sigma_{v}^2 > \sigma_{m}^2$, it follows that $1/\sigma_{v}^2 - 1/\sigma_{m}^2 < 0$, suggesting a negative relative bias for the RC method. In this case, the RC estimator exceeds the bias of the naive estimator if and only if
        \begin{equation*}
            \text{Relative Bias}_{\text{naive}} - \text{Relative Bias}_{\text{RC}} > 0.
        \end{equation*}
        However, based on 
        \begin{equation}\label{eq:pm}
            \begin{aligned}
                \text{Relative Bias}_{\text{naive}} - \text{Relative Bias}_{\text{RC}} &= \frac{C_1/\sigma^2 - C_1^2/\sigma^2 - 1/\sigma_{m}^2}{C_1^2/\sigma^2 + 1/\sigma_{m}^2} - \frac{1/\sigma_{v}^2 - 1/\sigma_{m}^2}{C_1^2/\sigma^2 + 1/\sigma_{m}^2}\\
                &= \frac{(C_1 - C_1^2)/\sigma^2 - 1/\sigma_{v}^2}{C_1^2/\sigma^2 + 1/\sigma_{m}^2},
            \end{aligned}
        \end{equation}
        $\sigma_{v}^2>0$, and $(C_1 - C_1^2)<0$ when $C_1 \in (-\infty,0]\cup [1,\infty)$, we have 
        \begin{equation*}
            (C_1 - C_1^2)/\sigma^2 - 1/\sigma_{v}^2 < 0, 
        \end{equation*}
        which implies that the RC estimator always achieves lower bias than the naive method under this scenario. 
    \end{enumerate}
    \item When $C_1 \in (\frac{1 - \sqrt{1 - 4\sigma^2/\sigma_{m}^2}}{2},\frac{1 + \sqrt{1 - 4\sigma^2/\sigma_{m}^2}}{2})$ and $\sigma^2 < \sigma_{m}^2/4$, the relative bias of the naive method is positive. As in the previous step, the relative bias of the RC estimator must be evaluated case by case.
    \begin{enumerate}
        \item When $\sigma_{v}^2 < \sigma_{m}^2$, the relative bias of the RC method is positive. Under this situation, the RC estimator has a larger bias than the naive estimator if and only if
        \begin{equation*}
            \text{Relative Bias}_{\text{naive}} - \text{Relative Bias}_{\text{RC}} < 0.
        \end{equation*}
        Based on equation \eqref{eq:pm}, we have
        \begin{equation*}
            \text{Relative Bias}_{\text{naive}} - \text{Relative Bias}_{\text{RC}}<0 \Leftrightarrow (C_1 - C_1^2)/\sigma^2 - 1/\sigma_{v}^2 < 0, 
        \end{equation*}
        which implies that
        \begin{equation*}
            \text{Relative Bias}_{\text{naive}} - \text{Relative Bias}_{\text{RC}}<0 \Leftrightarrow \sigma_{v}^2 < \frac{\sigma^2}{C_1 - C_1^2}.  
        \end{equation*}

        Therefore, when
        \begin{equation*}
            \sigma_{v}^2 < \min \{\sigma_{m}^2, \frac{\sigma^2}{C_1 - C_1^2}\}, 
        \end{equation*}
        the naive method exhibits smaller bias than the RC estimator. 

        \item When $\sigma_{v}^2 > \sigma_{m}^2$, the relative bias of the RC method is negative. Under this situation, the RC estimator has a larger bias than the naive estimator if and only if
        \begin{equation*}
            \text{Relative Bias}_{\text{naive}} + \text{Relative Bias}_{\text{RC}} < 0.
        \end{equation*}
        According to equation \eqref{eq:pp}, we have
        \begin{equation*}
            \text{Relative Bias}_{\text{naive}} + \text{Relative Bias}_{\text{RC}}<0 \Leftrightarrow (C_1 - C_1^2)/\sigma^2 + 1/\sigma_{v}^2 - 2/\sigma_{m}^2<0, 
        \end{equation*}
        which implies that
        \begin{equation*}
            \text{Relative Bias}_{\text{naive}} + \text{Relative Bias}_{\text{RC}}<0 \Leftrightarrow \sigma_{v}^2 > \frac{\sigma^2\sigma_{m}^2}{2\sigma^2 + (C_1^2 - C_1)\sigma_{m}^2} \text{ and }\sigma^2 > \frac{C_1 - C_1^2}{2}\sigma_{m}^2. 
        \end{equation*}

        Therefore, when
        \begin{equation*}
            \sigma_{v}^2 > \max \{\sigma_{m}^2, \frac{\sigma^2\sigma_{m}^2}{2\sigma^2 + (C_1^2 - C_1)\sigma_{m}^2}\}, \text{ and }\sigma^2 > \frac{C_1 - C_1^2}{2}\sigma_{m}^2,
        \end{equation*}
        the naive method exhibits lower bias. 
        
    \end{enumerate}
    
    \end{enumerate}

    Similarly, we can derive the conditions under which the absolute relative bias of the RC estimator is less than $R$, where $R\ge 0$. 

    \begin{enumerate}
        \item When $\sigma_{v}^2 < \sigma_{m}^2$, the relative bias of the RC method is positive, then
        \begin{equation*}
            \text{Relative Bias}_{\text{RC}} \le R \Leftrightarrow \frac{1/\sigma_{v}^2 - 1/\sigma_{m}^2}{C_1^2/\sigma^2 + 1/\sigma_{m}^2} \le R.
        \end{equation*}
        Thus,
        \begin{equation*}
        \frac{\sigma^2 \sigma_{m}^2}{\sigma^2 + RC_1^2\sigma_{m}^2 + R\sigma^2}\le \sigma_{v}^2 < \sigma_{m}^2.
        \end{equation*}
        \item When $\sigma_{v}^2 > \sigma_{m}^2$, the relative bias of the RC method is negative, then
        \begin{equation*}
            \text{Relative Bias}_{\text{RC}} \ge -R \Leftrightarrow \frac{1/\sigma_{v}^2 - 1/\sigma_{m}^2}{C_1^2/\sigma^2 + 1/\sigma_{m}^2} \ge -R.
        \end{equation*}
        Thus,
        \begin{equation*}
        \sigma_{m}^2 < \sigma_{v}^2 \le \frac{\sigma^2 \sigma_{m}^2}{\sigma^2 + RC_1^2\sigma_{m}^2 + R\sigma^2}, \text{ if }R\le \frac{\sigma^2}{C_1^2\sigma_{m}^2 + \sigma^2}.
        \end{equation*}
    \end{enumerate}

\end{document}